\documentclass[twocolumn,amssymb,showpacs,pre]{revtex4}
\usepackage{graphicx}
\begin{document}
\title{Ground state fidelity in bond-alternative Ising chains
with Dzyaloshinskii-Moriya interactions}

\author{Bo Li}
\email{libo.phy@cqu.edu.cn}
 \affiliation{Centre for Modern Physics and
Department of Physics, Chongqing University, Chongqing 400044, The
People's Republic of China}

\author{Sam Young Cho}
\email{sycho@cqu.edu.cn}
 \affiliation{Centre for Modern Physics and
Department of Physics, Chongqing University, Chongqing 400044, The
People's Republic of China}

\author{Hong-Lei Wang}
 \affiliation{Centre for Modern Physics and
Department of Physics, Chongqing University, Chongqing 400044, The
People's Republic of China}

\author{Bing-Quan Hu}
 \affiliation{Centre for Modern Physics and
Department of Physics, Chongqing University, Chongqing 400044, The
People's Republic of China}

\begin{abstract}
 A systematic analysis is performed for quantum phase transitions
 in a bond-alternative one-dimensional Ising model with a
 Dzyaloshinskii-Moriya (DM) interaction by using the fidelity of ground
 state wave functions based on the infinite matrix product states
 algorithm.
 For an antiferromagnetic phase, the fidelity per
 lattice site exhibits a bifurcation, which shows
 spontaneous symmetry breaking in the system.
 A critical DM interaction is inversely proportional to
 an alternating exchange coupling strength for a quantum phase transition.
 Further, a finite-entanglement scaling of von Neumann entropy
 with respect to truncation dimensions
 gives a central charge
 $c \simeq 0.5$ at the critical point.
\end{abstract}
\pacs{75.10.Pq, 64.70.Tg, 75.30.Kz}
% 75.10.Jm Quantized spin models
% 75.40.Mg Numerical simulation studies
% 75.30.Kz Magnetic phase boundaries (including magnetic transitions, meta-magnetism, etc.)
% $---------------------$

\maketitle

\section{Introduction}
 Recent advanced material technologies have made it possible to access
 low-dimensional quantum systems. Furthermore, material synthesis
 has offered a great opportunity to explore more intriguing
 lower-dimensional spin systems
 rather than well-understood conventional spin systems
 \cite{Verdaguer}.
 In such a low-dimensional system, for instance, alternating bond
 interactions and/or
 less symmetry interactions
 in spin lattices can be realizable in synthesizing two different magnetic atoms.
 Of particular importance, therefore, is
 understanding quantum phase transitions in which one-dimensional spin systems
 are unlikely found naturally.

 Normally, quantum fluctuations in a low-dimensional spin system
 are stronger than higher dimensional spin systems \cite{sachdev}.
 Quantum phase transitions driven by stronger quantum
 fluctuations then exhibit more interesting and novel quantum phenomena
 in low-dimensional spin systems.
 The effects of alternating bond
 interactions, especially,
 have been intensively studied theoretically
 in spin systems such as
 antiferromagnetic Heisenberg chains
  \cite{Kolezhuk,Yamamoto97,Dukelsky,Aplesnin,Onishi,Narumi},
 Heisenberg chains with next-nearest-neighbor bond alternations
 \cite{Capriotti,Maeshima},
 a tetrameric Heisenberg antiferromagnetic chain \cite{Gong},
 and two-leg spin ladders \cite{Fukui,Almeida}.
 A recent experiment has demonstrated a realization
 of a bond-alternating chain by
 applying magnetic fields in
 a spin-1/2 chain antiferromagnet \cite{Canevet}.

 In this study, we will consider one-dimensional Ising-type spin
 chains with an alternating exchange coupling.
 Actually, this bond alternation cannot destroy the
 antiferromagnetic phase of the uniform bond case but just quantitatively changes
 the ground state properties originating from a dimerization of the spin lattice.
 Then, a less symmetric interaction can play a significant role to induce a
 quantum phase transition. To see a quantum phase transition,
 we will employ a Dzyaloshinskii-Moriya (DM) interaction
 \cite{Moriya} which results from the spin-orbit coupling.

 Based on the ground state fidelity \cite{Zhou} with the iMPS presentation \cite{iTEBD},
 we discuss  the quantum criticality in the system.
 It is shown that a uniform DM interaction can destroy the antiferromagnetic
 phase, which is a continuous quantum phase transition,
 and its critical value is inversely proportional to
 the alternating exchange coupling strength.

\section{Model and numerical method}
 Let us start with a spin-1/2 Ising chain with antisymmetric
 anisotropic, and alternative bond interactions on the infinite-size lattice.
 Our system can be described by the spin Hamiltonian
 \begin{equation}
 H= \sum_{i=\infty}^\infty J_i S^{z}_{i}S^{z}_{i+1}
 + \vec{D}_{i}\cdot(\vec{S}_{i} \times \vec{S}_{i+1}), \label{Hamt}
 \end{equation}
 where
 $\vec{S}_i=(S^x_i, S^y_i, S^z_i)$ are the spin operators acting
 on the $i$-th site.
 The exchange interaction is chosen as $J_i =1-(-1)^i r$
 and the alternative bond interaction
 is characterized by the relative strength $r$ of exchange coupling
 for the even and odd lattice sites.
 To describe an antisymmetric anisotropic exchange coupling between
 the two spins on the lattice, we employ
 a uniform DM interaction $\vec{D}_i=\vec{D}$,
 that is characterized by
 the DM vector $\vec{D}=(D_x,D_y,D_z)$.
 For $r=0$ and $\vec{D}=0$,  Eq. (\ref{Hamt}) is reduced to the
 conventional Ising chain Hamiltonian.
 If $r=0$ and $\vec{D} = D \hat{z}$,
 Eq. (\ref{Hamt}) can be mapped onto the XXZ spin chain model
 which has a quantum phase transition from the gapped Neel or antiferromagnetic (AFM) phase
 to the gapless Luttinger Liquid (LL) phase at the critical point $D_c=1$ \cite{Soltani}.
 This study will then be focused on the antiferromagnetic exchange
 interaction $J_i \geq 0$, i.e., $0 \leqslant r \leqslant 1$, and
 a transverse DM interaction denoting $\vec{D}=(0, 0, D)$.

 The Hamiltonian in Eq. (\ref{Hamt}) is actually invariant
 under the transformation $U = \prod U_{2i}\otimes U_{2i+1}$
 with $U_{2i}=\sigma^x$ for $2i$-th site and
 $U_{2i+1}=\sigma^y$ for ($2i+1$)-th site.
 Our model Hamiltonian then possesses a $Z_2$ symmetry
 generated by the transformation $U$.
 The ground state of the system may undergo a spontaneous
 $Z_2$ symmetry breaking which gives rise to a quantum
 phase transition between an ordered phase and a
 disordered phase.

 For a quantum spin system with a finite $N$ lattice site,
 its wave function with the periodic boundary condition can be
 expressed in the matrix product state (MPS)
 representation \cite{Verstrate} as
  $|\Psi\rangle = \mathrm{Tr}\left[A^{[1]}A^{[2]} \cdots A^{[N]}\right]
  \, |s^{[1]}s^{[2]} \cdots s^{[N]}\rangle$, where
 $A^{[i]}$ is a site-dependent $\chi\times\chi$ matrix with the truncation
 dimension $\chi$ of the local Hilbert space at the $i$-th site,
 $|s^{[i]}\rangle$ is a basis of the local Hilbert space at the $i$-th
 site, and the physical index $s$ takes value $1,\cdots,d$ with the
 dimension $d$ of the local Hilbert space.
 This MPS representation for a finite lattice system can be extended to describe an infinite
 lattice system. To do this, for an infinite lattice,
 one may replace
 the matrix $A^{[i]}$ with $\Gamma^{[i]}\lambda^{[i]}$ \cite{iTEBD}, where
 $\Gamma^{[i]}$ is a three-index tensor and $\lambda^{[i]}$ is a
 diagonal matrix at the $i$-th site,
 which is called the \textit{canonical infinite matrix product state} (iMPS) representation.

 If system Hamiltonian is translational invariant for an infinite lattice, for instance,
 our system Hamiltonian describe by Eq. (\ref{Hamt}) has a two-site translational invariance,
 the two-site translational invariance allows us to reexpress the Hamiltonian
 as $H=\Sigma_i h^{[i,i+1]}$, where $h^{[i,i+1]}$ is
 the nearest-neighbor two-body Hamiltonian density.
 In such a case, one can introduce a two-site translational invariant iMPS representation,
 i.e., for the even (odd) sites A (B),
 only two three-index tensors $\Gamma_{A(B)}$
 and two diagonal matrices $\lambda_{A(B)}$
 can be applied in representing a system wave function:
 \begin{equation}
 |\Psi\rangle
 =\sum_{\{s^{[i]}\}}
 \cdots \Gamma_{A}\lambda_{A}\Gamma_{B}\lambda_{B}\Gamma_{A}
 \lambda_{A}\Gamma_{B}\lambda_{B} \cdots
 |\cdots s^{[i]}s^{[i+1]}s^{[i+2]}s^{[i+3]} \cdots \rangle. \label{wave}
 \end{equation}
 Note that, actually, for an infinite lattice sites,
 the diagonal elements of the matrix $\lambda_i$
 are the normalized Schmidt decomposition coefficients of the
 bipartition between the semi-infinite chains
 $L(-\infty,...,i)$ and $R(i+1,...,\infty)$.

 In order to find a ground state of our system in the iMPS
 representation, the infinite time-evolving block decimation
 (iTEBD) algorithm introduced by Vidal \cite{iTEBD} is employed.
 For a given initial state $|\Psi(0)\rangle$ and the Hamiltonian $H$,
 a ground-state wave function
 can be yield by the imaginary time evolution
 $|\Psi(\tau)\rangle = \exp\left[-H\tau\right]|\Psi(0)\rangle /|\exp(-H\tau)|\Psi(0)\rangle|$
 for a large enough $\tau$.
 To realize the imaginary time evolution operation numerically, the imaginary
 time $\tau$ is divided into the time slices $\delta\tau= \tau/N$
 and  a sequence of the time slice evolution gates approximates
 the continuous time evolution.
 Meanwhile,
 the time evolution operator $\exp\left[-h^{[i,i+1]} \delta\tau\right]$
 for $\delta\tau \ll 1$ is expanded to a product of the evolution operators acting
 on two successive $i$ and $i+1$ sites through the Suzuki-Trotter decomposition \cite{Suzuki}.
 After absorbing a time-slice evolution gate,
 in order to recover the iMPS representation,
 a singular value decomposition of a matrix is performed, which contracted from
 $\Gamma_A$, $\Gamma_B$, one $\lambda_A$ , two $\lambda_B$
 and the evolution operators. Then only the $\chi$ largest singular values are retained.
 This procedure yields the new tensors $\Gamma_A$, $\Gamma_B$ and $\lambda_A$
 that are used to update the tensors for all the sites.
 As a result, the translational invariance
 under two-site shifts is recovered.
 Repeating the above procedure until the ground-state energy
 converges yields the ground-state wave function of the system in the iMPS representation.

\section{Fidelity per lattice site and bifurcations}
 One can define a  fidelity $F(D,D')= | \langle\Psi(D')|
 \Psi(D)\rangle|$ from a ground state wavefunction $\Psi(D)$.
 A fidelity per lattice site (FLS) $d$ \cite{Zhou} can be
 defined as
\begin{equation}
\ln d(D,D') = \lim_{N \rightarrow \infty} \frac {\ln F(D,D')}{N}.
\label{dinf}
\end{equation}
 where $N$ is the system size. Remarkably, FLS is well
 defined and plays a similar role to an order parameter
 although $F(D,D')$ trivially becomes
 zero in the thermodynamic limit when $N$ approaches infinity.
 The FLS satisfies the properties inherited from
 fidelity $F(D,D')$:
 (i) normalization $d(D,D)=1$, (ii) symmetry $d(D,D')=d(D',D)$,
 and (iii) range $0 \le d(D,D')\le 1$.

%%%%%%%%%%%%%%%%%%%%%%%%%%%fig 1%%%%%%%%%%%%%%%%%%%%%%%%%%%%%%%%%%%%
\begin{figure}
\includegraphics [width=0.45\textwidth]{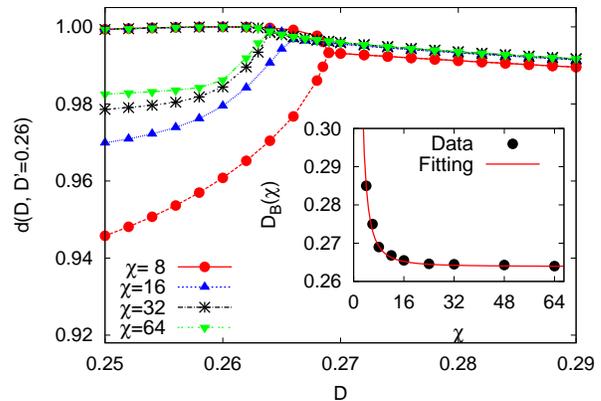}
\caption{ (Color online)
 Fidelity per lattice site $d(D,D'=0.26)$ as a function of
 $D$ for various truncation dimensions $\chi$ with $r=0.5$.
 For higher truncation dimension, a bifurcation point
 $D_B(\chi)$ moves to saturate to its critical value.
 Inset: Extrapolation of bifurcation point $D_B(\chi)$.
 From the numerical fitting function
 $D_B(\chi)=D_B(\infty)+ b\chi^{-c}$ with $b=0.335$ and
 $c=1.994$, the critical point is estimated as
 $D_c=D_B(\infty)=0.264$.}
 \label{FidelityBifur}
\end{figure}
%%%%%%%%%%%%%%%%%%%%%%%%%%%fig 1%%%%%%%%%%%%%%%%%%%%%%%%%%%%%%%%%%%%

 On adapting the transfer matrix approach \cite{iPEPS-fidelity},
 the iMPS representation of the ground-state wave-functions
 allows us to calculate the fidelity per lattice site (FLS)
 $d(D,D')$. Let us choose $|\Psi(D')\rangle$ as a reference
 state for the FLS $d(D,D')$.
 For $D'=0.26$, in Fig. \ref{FidelityBifur},
 the FLS $d(D,D')$ is plotted as a function of the DM
 interaction parameter $D$
 for various values of the truncation
 dimensions with a randomly chosen initial state
 in the iMPS representation.
 Figure \ref{FidelityBifur} shows a singular behavior of the FLS
 $d(D,0.26)$, which indicates that there occurs a quantum phase
 transition across the singular point.
 A bifurcation behavior of the FLS $d(D,0.26)$
 is also shown when the interaction parameter $D$ becomes smaller
 than its characteristic singular
 value that can be called a `bifurcation point' $D_B$.
 The bifurcation points depend on
 the truncation dimension $\chi$, i.e.,  $D_B=D_B(\chi)$.
 As the truncation dimension $\chi$ increases,
 the bifurcation occurs starting at the lower value of $D$.
 For $\chi \rightarrow \infty$,
 the bifurcation point $D_B(\infty)$ at which a bifurcation starts to occur
 can be extrapolated.
 In the inset of Fig. \ref{FidelityBifur},
 we use an extrapolation function $D_B(\chi)=a\,
 +\, b\, \chi^{-c}$, characterized by the coefficients $a$, $b$,
 and $c$ being a positive real
 number,
 which guarantees that
 $D_B(\infty)$ becomes a finite value.
 The numerical fitting gives $a = 0.264$, $b=0.335$, and $c=1.994$.
 From the extrapolation,
 the bifurcation point is shown to saturate to $D_B(\infty) \simeq a$ which
 can be regarded as a critical point $D_c = D_B(\infty)$ \cite{Bifur,Dai}.
 Actually, the critical point also corresponds
 to the pinch point of the FLS in the thermodynamics limit,
 i.e., $\chi \rightarrow \infty$.
 Consequently, a FLS bifurcation
 point $D_B(\chi)$ plays the role of a pseudo phase transition point for a given
 finite truncation dimension $\chi$ in the MPS representation.
 In addition, the continuous function behavior of the
 FLS across the bifurcation point implies
 that a continuous quantum phase transition occurs
 at the critical point \cite{Zhou}.

 In Fig.~\ref{FidelityBifur}, the bifurcation occurs for $D <
 D_B(\chi)$, which is captured in the iMPS representation with a
 randomly chosen initial state.
 In fact, the bifurcation behavior of the FLS means
 that there are two possible ground
 states for $D < D_B(\chi)$ while there is a single ground state for $D >
 D_B(\chi)$.
 Such a property of the ground states
 can be understood by the $Z_2$ symmetry of the Hamiltonian
 from the invariant transformation $U H U^\dagger=H$.
 In the thermodynamic limit, there are two possible ground states
 satisfying $U H U^\dagger=H$, that is,
 $\Psi_g$ or $U\Psi_g$.
 Once a spontaneous symmetry breaking happens,
 the system can choose one of two possible ground
 states  $\Psi_g$ or $U\Psi_g$, which indicate a broken symmetry
 phase.
 For symmetric phase, the system has a single ground state,
 being a linear combination of two possible ground state,
 which should satisfy the transformation invariance.
 In Fig.~\ref{FidelityBifur}, then, the FLS is plotted from
 the two fidelities, i.e., $F= |\langle \Psi_g(0.26) |
 \Psi_g (D)\rangle |$ (upper lines) and $F = |\langle \Psi_g(0.26) |
 U|\Psi_g (D) \rangle |$ (lower lines).
 Then, the bifurcation point is the transition point between
 the symmetry phase and the broken-symmetry phase.

\section{Phase diagram}
\begin{figure}
 \includegraphics[width=0.45\textwidth] {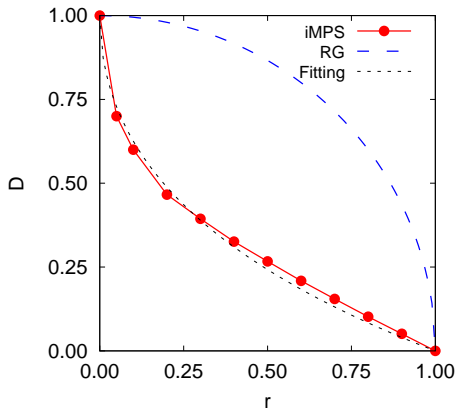}
 \caption{(Color online)
 Ground state phase diagram in the plane of the DM interaction and
 the alternating bond strengths.
 From the iMPS, the phase boundary shows that the $D_c$ is inversely
 proportional to the $\sqrt{r_c}$.
 A best fitting is denoted by the dotted line. This is in sharp contrast with  the results from the
 real space renormolaztion group (RSRG) method~\cite{XiangHao2010}.
 }
 \label{PhaseDigram}
\end{figure}

 From the iMPS with the numerical extrapolation of the bifurcation points,
 in Fig.~\ref{PhaseDigram},
 we draw the ground-state phase diagram in the interaction parameter
 $(r,D)$ plane.
 Below the phase boundary (red solid line) the system is in an antiferromagnetic
 phase while above the boundary
 the system is in a disordered phase.
 The phase diagram shows
 that the $D_c$ is inversely proportional to the $r_c$.
 A best fitting function (dotted line) of the critical points $(r_c,D_c)$
 is given by
 $D_c \approx (a(a+1))/(\sqrt{r_c}+a)-a$ with a single parameter $a=3.4$.
 The characteristic phase boundary can be understood as follows:
 If $r \neq 0$ and $D=0$,
 the lattice sites are dimerised and
 the Hamiltonian in Eq. (\ref{Hamt})
 has a two-site translational
 invariance due to the alternating bond coupling $r$.
 As assumed, for the antiferromagnetic exchange interaction $J_i > 0$, i.e.,
 for $0 \leq r \leq 1$, the Ising chain with the alternating
 bond coupling is in an antiferromagnetic state even though
 the lattice sites are strongly dimerised.
 If $D \neq 0$,
 the antisymmetric anisotropic DM interaction  can destroy the
 antiferromagnetic order originating from the antiferromagnetic
 exchange interaction $J_i$.
 The antiferromagnetic order
 may be destroyed more easily by the uniform
 antisymmetric anisotropic DM
 interaction for the dimerised lattice sites than for the Ising chain without
 the alternating bond coupling
 because the antiferromagnetic correlation
 between the sites becomes weaker due to the dimerised lattice sites.
 Then, to destroy the antiferromagnetic order,
 a stronger dimerisation of the lattice sites (bigger $r$)
 requires a much weaker uniform
 antisymmetric anisotropic interaction (much smaller $D$).
 Consequently,
 the phase boundary separating the antiferromagnetic phase
 and a disordered phase might have a inversely proportional relation
 between $D_c$ and $r_c$.

%%%%%%%%%%%%%%%%%%%%%%%%%%%fig 2%%%%%%%%%%%%%%%%%%%%%%%%%%%%%%%%
%%%%%%%%%%%%%%%%%%%%%%%%%%%%fig 2%%%%%%%%%%%%%%%%%%%%%%%%%%%%%%%%

 Very recently, a real space renormalization
 group (RSRG) approach \cite{XiangHao2010} has been applied in the
 same model and has shown that a symmetric phase boundary is given
 by $D_c\simeq \sqrt{1-r^2_c}$ (blue dashed line in Fig. \ref{PhaseDigram}).
 Compared with the RSRG approach,
 our results  from the iMPS show that there is a quite
 significant discrepancy in the phase boundary line because,
 contrast to the iMPS,
 the RSRG is an approximate method based on low-energy states, which implies that
 it does not capture properly a contribution from relevant higher energy states.
 Then, the ground state phase diagram from our fidelity approach based on
 the iMPS is more reliable and accurate.

\section{Quantum entanglement and phase transition}
 In order to understand more clearly the quantum phase transition
 in our system, let us consider quantum entanglement that
 can also detect a quantum phase transition \cite{ent_review}.
 To quantify the quantum entanglement,
 we employ the von Neumann entropy, which is a good measure of
 bipartite entanglement between two subsystems of a pure
 state \cite{Bennett}, because our ground states are in a pure state.
 Then, the spin chain can be partitioned into the two parts
 denoted by the left semi-infinite chain $L$  and the right semi-infinite chain $R$.
 The von Neumann entropy is defined as
 $S=-\mathrm{Tr}\varrho_L\log_2\varrho_L=-\mathrm{Tr}\varrho_R\log_2\varrho_R$
 in terms of the reduced density matrix $\varrho_L$ or $\rho_R$ of the
 subsystems $L$ and $R$.
 In the iMPS representation, the von Neumann entropy
 for the semi-infinite chains $L$ or $R$ becomes
\begin{equation}
 S_i=-\sum_{\alpha=1}^\chi \lambda_{i,\alpha}^2 \log_2 \lambda_{i,\alpha}^2,
 \label{entropy}
\end{equation}
 where $\lambda_{i,\alpha}$'s are diagonal elements of the matrix
 $\lambda$ that
 could be directly obtained in the iMPS algorithm.
 This is because, when one partitions the two semi-infinite
 chains $L(-\infty,\cdots,i)$ and $R(i+1, \cdots,\infty)$, one gets
 the Schmidt decomposition
 $|\Psi\rangle=\sum_{\alpha=1}^{\chi}\lambda_{i,\alpha} |\phi_L\rangle|\phi_R\rangle$.
 From the spectral decomposition, $\lambda_{i,\alpha}^2$ are actually
 eigenvalues of the reduced density matrices for the two
 semi-infinite chains $L$ and $R$. In our two-site translational
 invariant iMPS representation, there are two Schmidt coefficient
 matrices $\lambda_A$ and $\lambda_B$ that describe two possible ways
 of the partitions, i.e., one is on the odd sites, the other is on
 the even sites.
 From the $\lambda_A$ and $\lambda_B$,
 one can obtain the two von Neumann entropies depending on the odd- or even-site partitions.
%%%%%%%%%%%%%%%% fig 3%%%%%%%%%%%%%%%%%%%%%%%%%%%
\begin{figure}
 \includegraphics [width=0.45\textwidth]{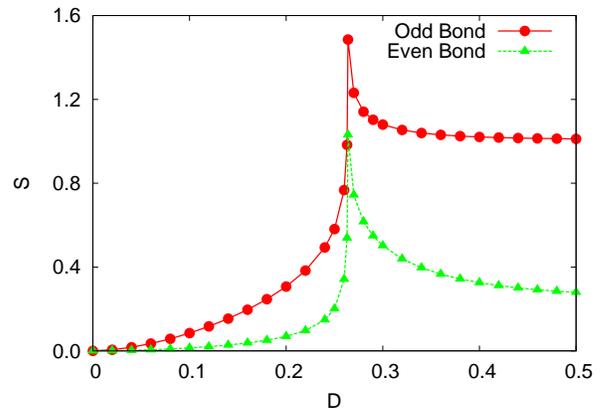}
 \caption{(Color online)
 The von Neumann entropy $S$ between left and right halves of a chain as
 a function of $D$ for $r=0.5$ and $\chi=32$.
 Both the von  Neumann entropies for odd and even bonds
 show a singularity at $D_c = 0.264$.}
 \label{Entanglement}
\end{figure}
%%%%%%%%%%%%%%%%%% fig 3%%%%%%%%%%%%%%%%%%%%%%%%%%%%

 In Fig.~\ref{Entanglement}, we plot
 the von Neumann entropy $S$
 as a function of the DM interaction strength $D$
 for even ($2i$ site) and odd (($2i+1$) site) bonds with $r=0.5$.
 The von Neumann entropy for the odd bonds
 is always larger than those for the even bonds
 because, by the definition,
 the odd-site exchange interaction $J_{2i}$
 is stronger that the even-site exchange interaction $J_{2i+1}$.
 Furthermore,
 it is shown that both the entropies for the even and odd bonds
 have a singularity at the same value of the DM interaction strength $D$.
 Note that the singularities of the entropies occur at
 the critical point $D_c=0.264$.
 This result then shows clearly that
 both the FLS $d$ and the von Neumann entropy $S$
 give the same phase transition point.
 As a consequence,
 in fact, the von Neumann entropy $S$ gives the same phase diagram
 from the FLS in Fig. \ref{PhaseDigram}.

 As discussed, for the antiferromagnetic state of our system,
 there are two possible ground states
 that are connected by the unitary transformation $U$.
 From the bifurcation of FLS, then, one might expect
 a bifurcation in the von Neumann entropy too.
 However, contrary to the FLS $d$, in Fig.~\ref{Entanglement},
 no bifurcation is seen in the von Neumann entropy $S$ for
 the antiferromagnetic state even though the initial state
 is randomly chosen in the iMPS representation.
 The reason for the absence of bifurcation in the von Neumann entropy
 is why the singular values $\lambda_i$ in Eq.~(\ref{entropy})
 do not depend on the unitary transformation because the
 unitary transformation $U$ acts only on a single site of
 the spin lattice in the iMPS representation.

\section{Central charge and universality class}
 At a critical point, characteristic singular behaviors
 of thermodynamics system properties depend only on few features
 such as dimensionality and symmetry, which can be classified by the
 concept of universality classes.
 Especially, the central charge can be used for the classification
 of universality classes \cite{Cardy,Calabrese}.
 Owing to implement the iMPS representation,
 we can obtain a central charge $c$ and a so-called
 finite-entanglement scaling exponent $\kappa$ numerically
 via the unique behaviors
 of the correlation length $\xi$ and the von Neumann entropy $S$
 with respect to the truncation dimension $\chi$ at a critical point
 \cite{Corrlen,GVidal}, i.e.,
\begin{eqnarray}
  \xi & = & a\chi^\kappa, \\
  S & = & \frac{c\kappa}{6}\log_2{\chi}.
\end{eqnarray}

%%%%%%%%%%%%%%%%%%%%%%%%%%%%%%fig 4%%%%%%%%%%%%%%%%%%%%%%%%%%%%%%%
\begin{figure}
  \includegraphics[width=0.45\textwidth]{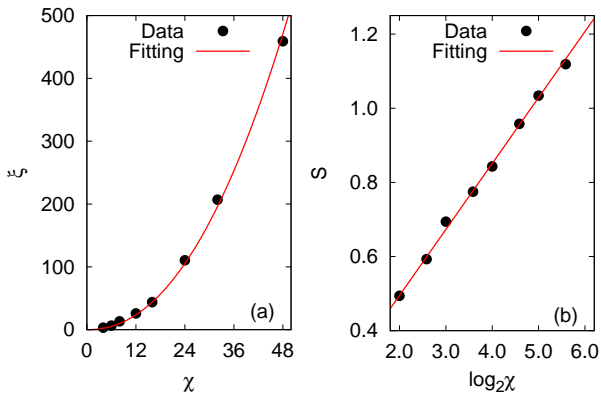}
  \caption{\label{Fig4}
  (Color online) (a) Correlation length $\xi$ as a function of the
  truncation dimension $\chi$ at the critical point. The power curve fitting
  $\xi=a \chi^ \kappa$ yields $a=0.112$ and $\kappa=2.132$.
  (b) Scaling of the von Neumann entropy $S$ with the truncation
  dimension $\chi$ at the critical point.
  For $\kappa=2.132$ from (a), the linear fitting
  $S=(c\kappa/6)\log_2{\chi}+b$ yields the central charge
  $c\approx 0.494$.
  Here, the alternating bond strength is chosen as $r=0.5$.
  }
\end{figure}
%%%%%%%%%%%%%%%%%%%%%%%%%%%%%fig 5%%%%%%%%%%%%%%%%%%%%%%%%%%%%%%%

 In Fig.~\ref{Fig4}, the correlation length $\xi$
 and the von Neumann entropy $S$ as a function of the truncation dimension
 $\chi$ at the critical point $D_c$ for $r=0.5$.
 Here, the truncation dimensions are taken as $\chi= 4, 6, 8, 12, 16, 24, 32$, and $48$.
 It is shown that both the correlation length $\xi$
 and the von Neumann entropy $S$ diverges as the truncation
 dimension $\chi$ increases.
 From a power-law fitting on the correlation length $\xi$,
 we have $\kappa = 2.132$ and $a = 0.112$.
 As shown in Fig. \ref{Fig4} (b),
 our numerical result demonstrates a linear scaling behavior,
 which gives a central charge $c \simeq 0.494$ with $\kappa = 2.132$.
 Our central charge is close to
 the exact value $c = 1/2$. Consequently, the quantum phase transition in
 our system
 is in the same universality class as the quantum
 transverse field Ising model.

\section{Summary}
 Quantum phase transitions have been investigated in the Ising
 chain with the Dzyaloshinskii-Moriya interaction as well as the
 alternating bond-coupling.
 The FLS and its bifurcation have clearly shown a characteristic singular point
 as a signature of the quantum phase transition and
 behaves as a continuous function, which shows a continuous phase transition
 occurring at the critical point.
 The phase diagram was obtained from the FLS and the von Neumann
 entropy.
 With a finite-entanglement scaling of the von Neumann entropy
 with respect to the truncation dimension in the iMPS representation,
 a central charge was estimated to be $c \simeq 0.5$,
 which shows that
 the system is in the same universality class with the quantum transverse
 field Ising model.

\begin{acknowledgments}
 We thank Huan-Qiang Zhou for helpful discussions.
 This work was supported by the Fundamental Research Funds for the
 Central Universities (Project No. CDJZR10100027). SYC
 acknowledges the support from the NSFC under Grant No.10874252.
\end{acknowledgments}


\begin{thebibliography}{99}
 \bibitem{Verdaguer}
 Mallah T, Thiebaut S, Verdaguer M and Veillet P 1993 {\it Science} {\bf 262} 1554 \\
 Sato O, Lyoda T, Fujishima A and Hashimoto K 1996 {\it Science} {\bf 272} 704 \\
 Verdaguer M, Gleizes A, Renard J P and Seiden J 1984 {\it Phys. Rev.} B {\bf 29} 5144 \\
 Kahn O, Pei Y, Verdaguer M, Renard J P and Sletten J 1988 {\it J. Am. Chem. Soc.} {\bf 110} 782 \\
 Koningsbruggen P J van, Kahn O, Nakatani K, Pei Y, Renard J P, Drillon M and Leggol P 1990 {\it Inorg. Chem.} {\bf 29} 3325 \\
 Zheludev A, Maslov S, Yokoo T, Raymond S, Nagler S E and Akimitsu J 2001 {\it J. Phys.: Condes. Matter} {\bf 13} R525 \\
 Hagiwara M, Minami K, Narumi Y, Tatani K and Kindo K 1998 {\it J. Phys. Soc. Jpn.} {\bf 67} 2209 \\
 Yamamoto S, 2000 {\it Phys. Rev.} B {\bf 61} R842 \\
 Culp J T, Park J H, Meisel M W and Talham D R 2003 {\it Inorg. Chem.} {\bf 42} 2842

\bibitem{sachdev} Sachdev S 1999 {\it Quantum Phase Transitions}, 2nd Edition.
  (Cambridge University Press, Cambridge)

 %%%%%%%%bond alternating %%%%%%%%%%%%%%%%%%%%%%%%%%%%
 \bibitem{Kolezhuk} Kolezhuk A K 1996 {\it Phys. Rev.} B {\bf 53} 318


 \bibitem{Yamamoto97} Yamamoto S 1997 {\it Phys. Rev.} B {\bf 55} 3603

 \bibitem{Dukelsky} Dukelsky J and Pittel S 1997 {\it Phys. Rev.} B {\bf 56} 10770

 \bibitem{Aplesnin} Aplesnin S S 2000 {\it Phys. Rev.} B {\bf 61} 6780

 \bibitem{Onishi} Onishi H and Miyashita S 2001 {\it Phys. Rev.} B {\bf 64} 014405

 \bibitem{Narumi} Narumi Y, Hagiwara M, Kohno M and Kindo K 2001 {\it Phys. Rev. Lett.} {\bf 86} 324

 \bibitem{Capriotti} Capriotti L, Becca F, Sorella S and Parola A 2003 {\it Phys. Rev.} B {\bf 67} 172404

 \bibitem{Maeshima} Maeshima N, Okunishi K, Okamoto K and Sakai T 2004 {\it Phys. Rev. Lett.} {\bf 93} 127203

 \bibitem{Gong} Gong S-S and Su G 2008 Phys. Rev. B {\bf 78}, 104416

 \bibitem{Fukui} Fukui T and Kawakami N 1998 {\it Phys. Rev.} B {\bf 57} 398

 \bibitem{Almeida} Almeida J, Martin-Delgado M A and Sierra G 2007 {\it Phys. Rev.} B {\bf 76} 184428

%%%%%%%%%%%%%%%%%%%%%%%%%%%%%%%%%%%%%%%%%%%%%%%%%%

 \bibitem{Canevet}
 Can\'evet E, Grenier B, Yoshida Y, Sakai N, Regnault L-P, Goto T, Fujii Y and Kawae T 2010 {\it Phys. Rev.} B {\bf 82} 132404

 \bibitem{Moriya} Dzyaloshinsky I 1958 {\it J. Phys. Chem. Solids} {\bf 4} 241 \\
 Moriya T 1960 {\it Phys. Rev.} {\bf 120} 91

 \bibitem{Zhou} Zhou H-Q and J.P. Barjaktarevi$\check{\rm c}$ J P 2008 {\it J. Phys. A: Math. Theor.} {\bf 41} 412001 \\
     Zhou H-Q, Zhao J-H and Li B 2008 {\it J. Phys. A: Math. Theor.} {\bf 41} 492002 \\
     Zhou H-Q arXiv:0704.2945

 \bibitem{iTEBD} Vidal G 2007 {\it Phys. Rev. Lett.} {\bf 98} 070201\\
   Vidal G 2004 {\it Phys. Rev. Lett.} {\bf 93} 040502 \\
   Vidal G 2003 {\it Phys. Rev. Lett.} {\bf 91} 147902

\bibitem{Verstrate} Verstraete F, Porras D and Cirac J I 2004 {\it Phys. Rev. Lett.} {\bf 93} 227205

 \bibitem{Suzuki} Suzuki M 1990 {\it Phys. Lett. A} {\bf 146} 319

 \bibitem{Soltani} Perk J H H and Capel H W 1976 {\it Phys. Lett.} A {\bf 58} 115 \\
  Jafari R, Kargarian M, Langari A and Siahatgar M 2008 {\it Phys. Rev.} B {\bf 78} 214414 \\
  Kadar Z and Zimboras Z 2010 {\it Phys. Rev.} A {\bf 82} 032334 \\
  Soltani M R, Mahdavifar S, Akbari A and Masoudi A A 2010 {\it J. Supercond. Nov. Magn.} {\bf 23} 1369

 \bibitem{iPEPS-fidelity} Zhou H-Q, Or\'us R and G. Vidal 2008 {\it Phys. Rev. Lett.} {\bf 100} 080601

 \bibitem{Bifur} Zhao J-H, Wang H-L, Li B and Zhou H-Q, 2010 {\it Phys. Rev.} E {\bf 82} 061127

 \bibitem{Dai} Dai Y-W, Hu B-Q, Zhao J-H and Zhou H-Q, 2010 {\it J. Phys. A: Math. Theor.} {\bf 43}  372001

 \bibitem{XiangHao2010} Hao X 2010 {\it Phys. Rev.} A {\bf 81} 044301

 \bibitem{ent_review} Amico L, Fazio R, Osterloh A and Vedral V 2008 {\it Rev. Mod. Phys.} {\bf 80} 517

 \bibitem{Bennett} Bennett C H, Bernstein H J, Popescu S and Schumacher B 1996 {\it Phys. Rev.} A {\bf 53} 2046

 \bibitem{Cardy} Cardy J, 1996 {\it Scaling and Renormalization in Statistical Physics}. (University of Oxford)


 \bibitem{Calabrese}
 Calabrese P and Cardy J 2009 {\it J. Phys. A} {\bf 42} 504005


 \bibitem{Corrlen} Tagliacozzo L, Oliveira Thiago R de, Iblisdir S and  Latorre J I 2008 \emph{Phys. Rev.} B \textbf{78} 024410 \\
     Pollmann F, Mukerjee S, Turner A and Moore J E 2009 {\it Phys. Rev. Lett.} {\bf 102} 255701

 \bibitem{GVidal}
 Vidal G, Latorre J I, Rico E and Kitaev A 2003 {\it Phys. Rev. Lett.} {\bf 90} 227902
\end{thebibliography}
\end{document}